# The protoplanetary disc HD 163296 as observed by ALMA

P.N. Diep*, D.T. Hoai, N.B. Ngoc, P.T. Nhung, N.T. Phuong, T.T. Thai, and P. Tuan-Anh

*Department of Astrophysics, Vietnam National Space Center, Vietnam Academy of Science and Technology*

**Abstract:**

HD 163296 is one of the few protoplanetary discs displaying rings in the dust component. The present work uses ALMA observations of the 0.9 mm continuum emission, which have significantly better spatial resolution (~8 au) than previously available, to provide new insight into the morphology of the dust disc and its double ring structure. The disc is shown to be thin, and its position angle and inclination with respect to the sky plane are accurately measured as are the locations and shapes that characterise the observed ring-gap structure. Significant modulation of the intensity of the outer ring emission is revealed and discussed. In addition, earlier ALMA observations of the emission of three molecular lines, CO(2-1), $C^{18}O$(2-1), and $DCO^+$(3-2) with a resolution of ~70 au are used to demonstrate the Keplerian motion of the gas, which is found to be consistent with a central mass of 2.3 solar masses. An upper limit of ~9% of the rotation velocity is placed on the in-fall velocity. The beam size is shown to give the dominant contribution to the line widths, accounting for both their absolute values and their dependence on the distance to the central star.

**Keywords:** planetary systems, protoplanetary discs, submillimetre astronomy.

**Classification number:** 2.1

## Introduction

The study of ring-like structures observed in the dust emission of protoplanetary discs is expected to shed light on the mechanisms governing the formation of planets. Only two such discs were known before the discovery of the ring-gap structure of the HD 163296 disc [1]: TW Hya [2] and HL Tau [3]. Although still being debated, such ring-gap structure is thought to be associated with the presence of newly formed giant planets [4]. It is, therefore, important to identify effects that differentiate planet formation from other gap-opening mechanisms, such as aggregation of solids in low turbulence regions [5] and changes in dust opacity at the frost line of volatile elements [6, 7]. Keplerian shear resulting from the radial velocity gradient can cause turbulence, which, at a certain distance from the star, can lead to the formation of a gap similar to those carved by planets. A difference between this and planet formation is that a planet sucks up all the material (both gas and dust) around it, whereas turbulence removes the dust but not the gas.

HD 163296, the third system known to host multiple rings in dust emission at millimetre wavelength, is a Herbig Ae star of intermediate mass (2.3 solar masses). It is about 4 million years old and is located at a distance of 122 pc from Earth [8]. The gas disc is in Keplerian motion and has a radius of about 550 au, and the millimetre continuum emission from solid particles is confined to within 250 au of the star [9, 10].

Both gas and dust emissions from HD 163296 have been observed by ALMA. According to the analysis of Isella, et al. [1], the 1.3 mm continuum emission observed with a spatial resolution of 25 au reveals three concentric gap-ring pairs, with the dust-depleted gaps located at ~54, ~100, and 160 au from the central star. The gas morphology displayed by CO(2-1), $^{13}$CO(2-1), and $C^{18}O$(2-1) emissions shows no clear evidence for ring-gap modulation, although small deviations from a smooth radial dependence have

*Corresponding author: Email: pndiep@vnsc.org.vn*

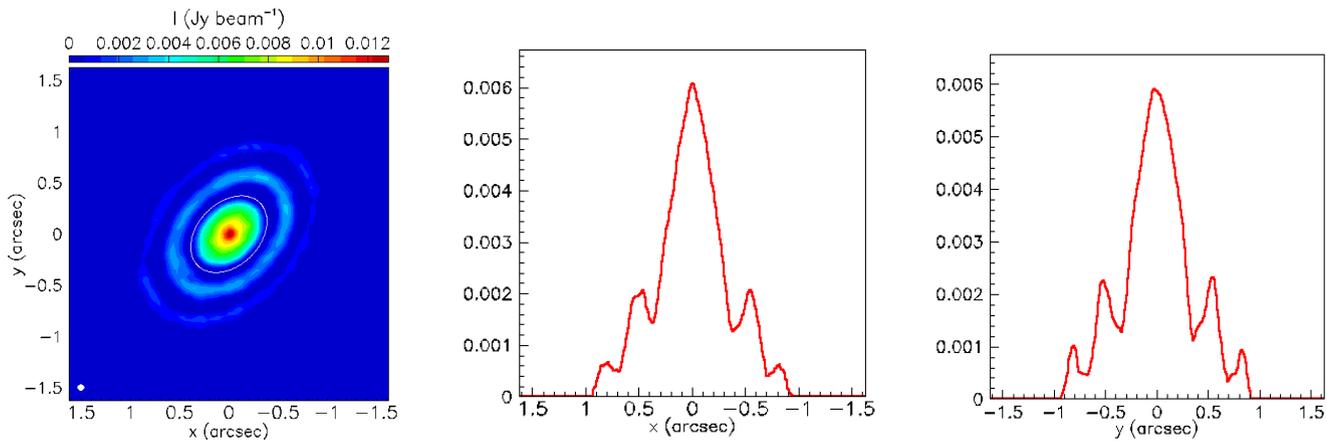

**Fig. 1.** Left: sky map of the continuum intensity. The white ellipse shows the region used to characterise the geometry of the disc. The beam is shown in the lower left corner of the map. Centre and right: projections on the $x$ and $y$ axes integrated over $y$ and $x$ respectively (in units of Jy beam$^{-1}$ arcsec).

been observed with the same resolution of ~25 au. Isella, et al. [1] interpret these as the result of a gas deficit restricted to the outer dust gaps but absent from the inner gap. They suggest that the outer gaps, at radial distances of 100 au and 160 au, are created by planets, probably about the mass of Saturn, but that the inner gap is due to gas turbulence or other physical phenomena within the disc. However, they do not exclude other possible interpretations. They evaluate the orbital radius and mass of the hypothetical planets from the location and shape of the dust gaps. The present study aims to provide additional information about this very interesting protoplanetary system.

### Continuum emission

*Observations*

The continuum data used in the present study were collected on August 15$^{th}$, 2017 and were reduced by the ALMA staff. The beam size is 0.069×0.061 arcsec$^2$ (FWHM) with a position angle of -88.8°, three times smaller than was found in earlier observations [1]. The data have been corrected for the proper motion of the source, (-7.61, -39.42) mas yr$^{-1}$, which implies that the source has moved (-0.14, -0.70) arcsec to the south-west direction from its position in J2000. The continuum, observed at 330.588 GHz (~0.9 mm wavelength), displays a well-behaved Gaussian noise with a standard deviation of 0.26 mJy beam$^{-1}$. In what follows, unless otherwise stated, we apply a 3-σ cut to the data.

*Sky plane morphology*

Figure 1 (left) shows the map of the continuum intensity. The abscissa $x$ and ordinate $y$ measure offsets in arcsec from the central star with $x$ pointing east and $y$ pointing north. The middle and right panels show the projections of intensity on the $x$ and $y$ axes. The dust continuum emission is found to display central symmetry and to extend up to ~±1 arcsec on the sky plane.

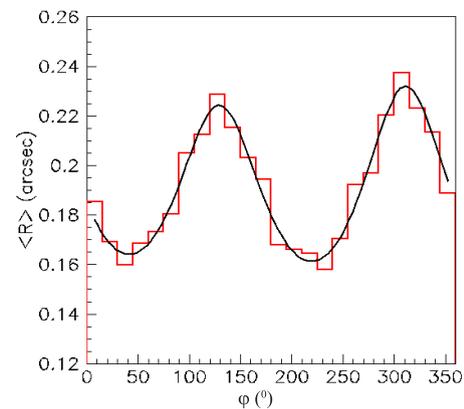

**Fig. 2.** Dependence of <$R$> on position angle $\varphi$.

Figure 2 displays the dependence on $\varphi$ of the mean value of the projected distance to the central star, <$R$>, where $\varphi=90°-tan^{-1}(y/x)$ is the position angle measured counter-clockwise from north, and $R=\sqrt{(x^2+y^2)}$. <$R$> is calculated using intensity as weight over the region contained inside the white ellipse shown in Fig. 1 (left), defined as having a semi-major axis of 0.44 arcsec, a semi-minor axis of 0.31 arcsec, and a position angle of the major axis of 138°. It is small enough to avoid contribution of the bright ring at larger distance from the star and large enough not to introduce any bias in the evaluation of the geometry parameters of the disc. A fit to an ellipse of the dependence of <$R$> on $\varphi$ gives position angle of the major axis $\varphi_0$, semi-major axis $a$, and semi-minor axis $b$ of 130°, 0.23 arcsec, and 0.16 arcsec, respectively. When interpreted as a thin and flat circular disc inclined by an angle $\theta$ with respect to the plane of the sky, $\theta=\cos^{-1}(b/a)=44^0$. Isella, et al. [1] quote values of $132^0$ for $\varphi_0$ and $42^0$ for $\theta$, which are, respectively, $2^0$ larger and $2^0$ smaller than the present evaluations. This indicates good agreement since systematic uncertainties are expected to be at that level.

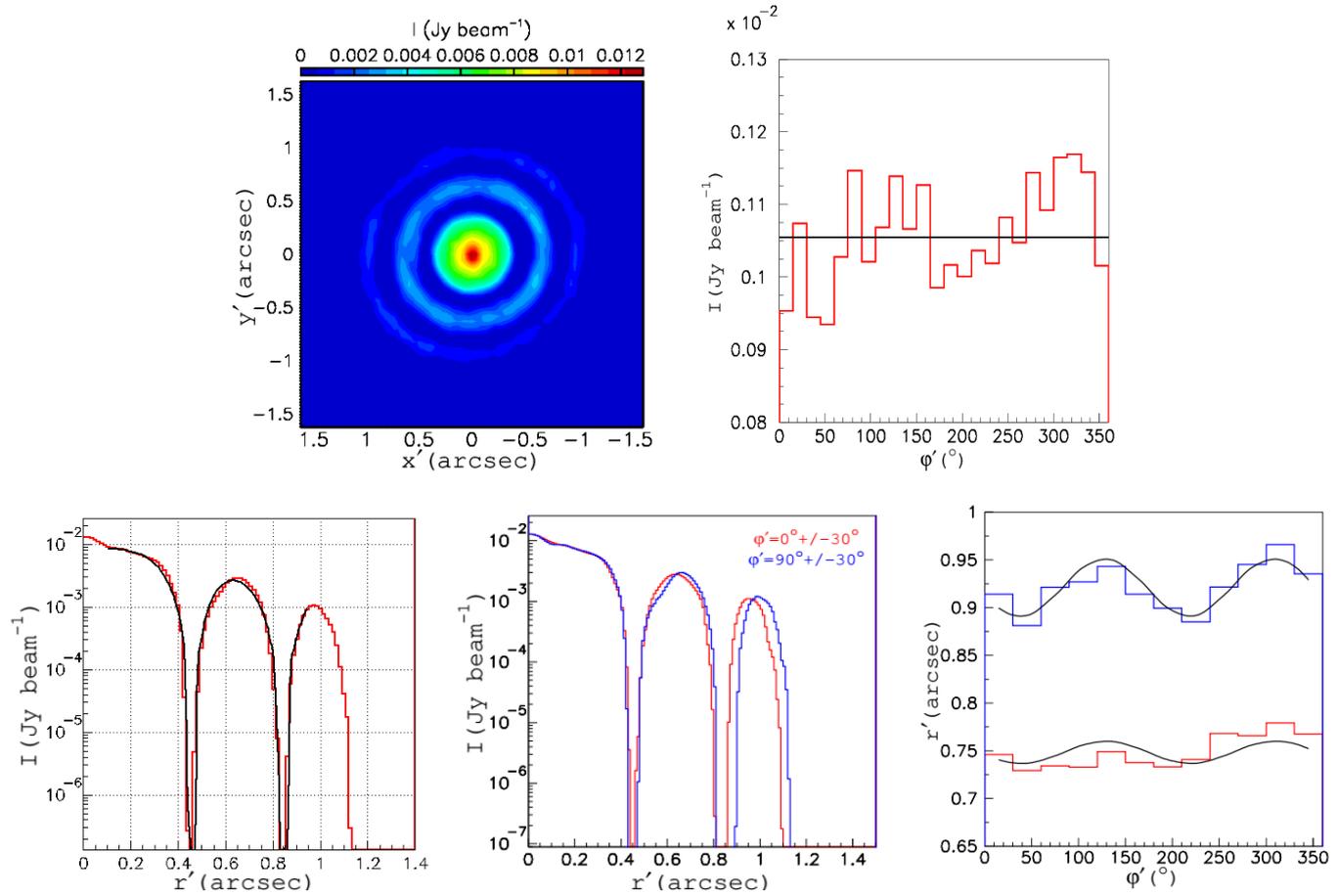

**Fig. 3.** Top left: intensity map de-projected in the disc plane ($x'$, $y'$). Top right: dependence on position angle $\varphi'$ measured in the disc plane of the intensity averaged over $r'<1.31$ arcsec; note the off-set scale of ordinate. Bottom left: dependence on $r'$ measured in the disc plane of the intensity averaged over $\varphi'$; the black curves are Gaussian fits to the gaps. Bottom centre: this is the same as the graph on the lower left for position angles $\varphi'$ less than $30^0$ away from the minor axis (red) and less than $30^0$ away from the major axis (blue). Bottom right: dependence on $\varphi'$ of the radius of the inner (red) and outer (blue) edges of the outer gap; the curves are fits of the form $r'=r'_0+k\cos2(\varphi'-\varphi'_0)$.

### De-projected morphology

Figure 3 (top left) shows the de-projected intensity map in the disc plane. What is meant here by de-projection is a simple transformation from $(x,y)$ coordinates to $(x',y')$ coordinates, which is defined as follows:

$x'=x\cos40^0-y\sin40^0$

$y'=(x\sin40^0+y\cos40^0)/\cos44^0$

Such a transformation would be accurate if the disc was perfectly flat and thin. In the disc plane, we define position angle $\varphi'$ and radial distance $r'$ from $x'$ and $y'$ in the same way as $\varphi$ and $R$ were defined from $x$ and $y$: $\varphi'=90^0-\tan^{-1}(y'/x')$ and $r'=\sqrt{(x'^2+y'^2)}$.

Figure 3 (top right) shows the dependence on $\varphi'$ of the intensity averaged over $r'<1.31$ arcsec. It displays small fluctuations having a standard deviation of only 7% of the mean value (1.1 mJy beam$^{-1}$). Fig. 3 (bottom left) displays the dependence on $r'$ of the intensity averaged over $\varphi'$. Compared with the results of Isella, et al. [1], the improved resolution produces much deeper gaps between the rings but provides no evidence for a third gap at $r'\sim1.31$ arcsec. This is confirmed in Fig. 4 where the 3-$\sigma$ cut has been lifted. Gaussian fits to the gaps and rings give values of the mean radius and standard deviation. These are listed in Table 1 and are in excellent agreement with the mean radii of the gaps quoted by Isella, et al. [1] as 0.44 arcsec and 0.81 arcsec, respectively.

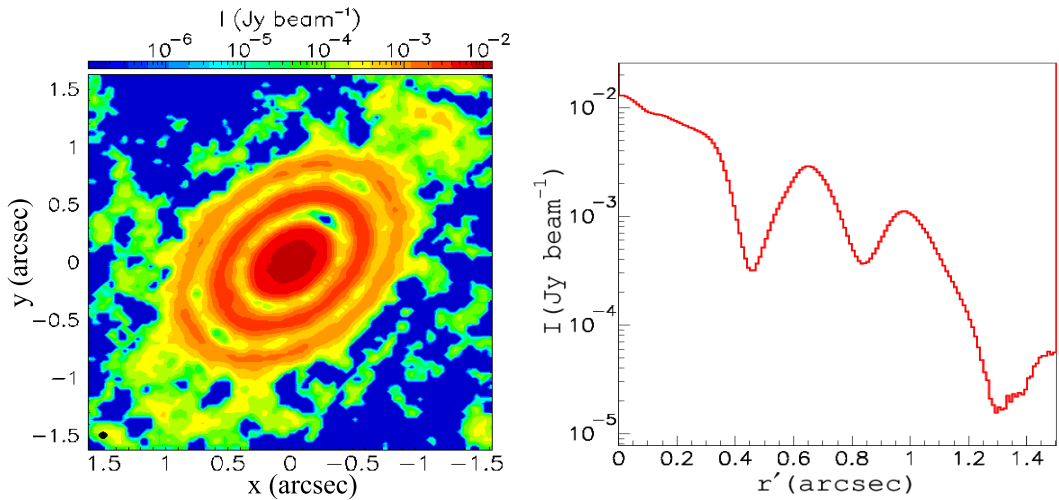

**Fig. 4. Continuum intensity without application of the 3-σ cut.** Left: map on the sky plane. Right: dependence on $r'$ (averaged on $\varphi'$).

**Table 1. Parameters defining the gap-ring structure of the dust disc.**

| Distance from central star (arcsec), Gaussian fits to rings and gaps | | | | | | | |
|---|---|---|---|---|---|---|---|
| **Inner ring** | | **Outer ring** | | **Inner gap** | | **Outer gap** | |
| Mean | rms | Mean | rms | Mean | rms | Mean | rms |
| 0.65 | 0.07 | 0.98 | 0.05 | 0.44 | 0.11 | 0.82 | 0.13 |
| **De-projection (see text)** | | | | | | | |
| *Disc* | | *Outer gap (inner edge)* | | *Outer gap (outer edge)* | | | |
| PA | $\theta$ | $r'_0$ | $\theta$ | $r'_0$ | $\theta$ | | |
| 40° | 44° | 0.75 | 44° | 0.92 | 46° | | |

If the disc was thick, the edges of the gaps would be smeared along the $y'$ axis but not along the $x'$ axis, typically by a quantity at the scale of the disc thickness. Indeed, the standard deviations of the gap Gaussians ~0.12 arcsec are upper limits to such smearing. To check this, we compare in Fig. 3 (bottom centre) the radial distribution of the de-projected intensity in two 60° wide wedges bracketing the minor and major axes, defined as $|\sin\varphi'|<1/2$ and $|\cos\varphi'|<1/2$, respectively. The absence of a significant difference for the inner gap shows that the thickness of the dust disc cannot significantly exceed 10 au at $r' \sim 80$ au. However, the outer gap, while showing no significant $\varphi'$ dependence of the smearing of its edges, shows a clear dependence on $\varphi'$ of both gap width and position. This is further illustrated in Fig. 3 (bottom right), which displays the dependence on $\varphi'$ of the position of the edges of the outer gap; the outer edge displays significant deviation from a circle. A fit to an ellipse of the form $r'=r'_0+k\cos2(\varphi'-\varphi'_0)$ would suggest an inclination angle ($\theta=44°-k/r'_0$) of 46°, differing slightly from that of the disc plane. However, it might also be that the assumption that the disc is thin and flat, which was used to evaluate the inclination, does not apply to the outer ring, in which case the deviation from a circle could be associated with a dependence on $\varphi'$ of the intensity or, more generally, of the morphology of the outer ring. Indeed, the dependence on $\varphi'$ of the intensity, averaged over $0.44<r'<0.82$ arcsec for the inner ring and over $0.82<r'<1.31$ arcsec for the outer ring, which is illustrated in Fig. 5, shows that while the inner disc displays small fluctuations of only ~3%, the outer ring displays much larger fluctuations of ~31%. Therefore, it is unjustified to interpret the fluctuations of the mean radius of the outer disc as resulting from a different inclination of its plane.

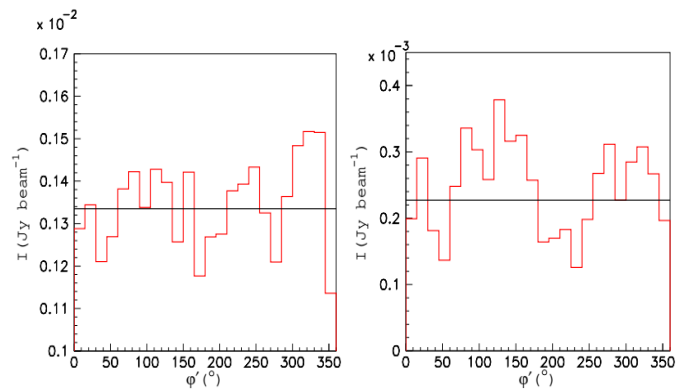

**Fig. 5. Dependence on $\varphi'$ of the average intensity measured in the inner ring (0.44<$r'$<0.82 arcsec, left panel) and in the outer ring (0.82<$r'$<1.31 arcsec, right panel).** Note the different scales of ordinate.

## Line emissions

### Observations

To study the gas present in the HD 163296 disc, we use ALMA archival observations made on June 4$^{th}$, 2014 for project 2013.1.00366.S. Their spatial resolution, ~0.7 arcsec (85 au), is the best currently available, since the data used by Isella, et al. [1], with a spatial resolution of 25 au, are inaccessible to the public. Three line emissions, CO(2-1), C$^{18}$O(2-1), and DCO$^{+}$(3-2), were observed in Cycle 2 of ALMA operation with spectral resolution of 0.04 km s$^{-1}$. The data were reduced by the ALMA staff into datacubes consisting of 300×300 pixels, each measuring 0.1×0.1 arcsec$^2$ with 800 bins of velocity, each 0.04 km s$^{-1}$ wide. Table 2 below lists the main parameters of relevance. The noise distribution is well-behaved in each of the three lines. In what follows, unless otherwise stated, we apply a 3-$\sigma$ cut to each datacube element, as was done for the study of continuum emission.

**Table 2. Information of relevance to the line emissions.**

| Line | CO(2-1) | C$^{18}$O(2-1) | DCO$^{+}$(3-2) |
|---|---|---|---|
| Beam Size (arcsec$^2$) | 0.69×0.53 | 0.70×0.55 | 0.70×0.57 |
| Beam Position Angle | −80.9$^0$ | −80.3$^0$ | −79.4$^0$ |
| Noise ($\sigma$, mJy beam$^{-1}$) | 5.6 | 3.6 | 3.0 |
| Right Ascension (J2000) | 17h 56m 21.2 s | | |
| Declination (J2000) | -21$^0$ 57' 22'' | | |
| Systemic Velocity | 5.80 km s$^{-1}$ | | |
| Pixel Size | 0.1×0.1 arcsec$^2$ | | |
| Velocity Bin Size | 0.04 km s$^{-1}$ | | |

### Main features

In what follows, we normally limit the analysis to ellipses where the signal is well above noise. These ellipses have semi-major and minor axes of (*a*, *b*)=(4.8, 3.8) arcsec, (3.5, 2.4) arcsec and (3.3, 2.4) arcsec for CO(2-1), C$^{18}$O(2-1), and DCO$^{+}$(3-2), respectively, with a common position angle of 135$^o$. Fig. 6 displays the Doppler velocity spectra of CO(2-1), C$^{18}$O(2-1), and DCO$^{+}$(3-2) emissions integrated over these ellipses. They are symmetric about the systemic velocity of 5.80 km s$^{-1}$ taken as origin. We limit the present analysis to Doppler velocities not exceeding 6 km s$^{-1}$ in absolute value.

Figure 7 displays sky maps of the integrated intensity and of the mean value and standard deviation of the Doppler velocity spectra. The mean Doppler velocity is seen to display a characteristic rotation pattern about the disc axis while the rms deviation from the mean is nearly isotropic. As was observed earlier by Isella, et al. [1] the intensity maps show no sign of a ring-gap structure, but their common orientation and aspect ratio are, at least qualitatively, the same as observed in continuum emission. Accordingly, in what follows, we use the same disc plane coordinates (*x'*,*y'*) as we used in the study of continuum emission. However, as the gas disc is expected to be significantly thicker than the dust disc, the thin disc hypothesis is no longer valid and the transformation of coordinates from (*x*,*y*) to (*x'*,*y'*) should no longer be strictly identified with de-projection as the disc thickness may cause significant smearing along the *y'* axis.

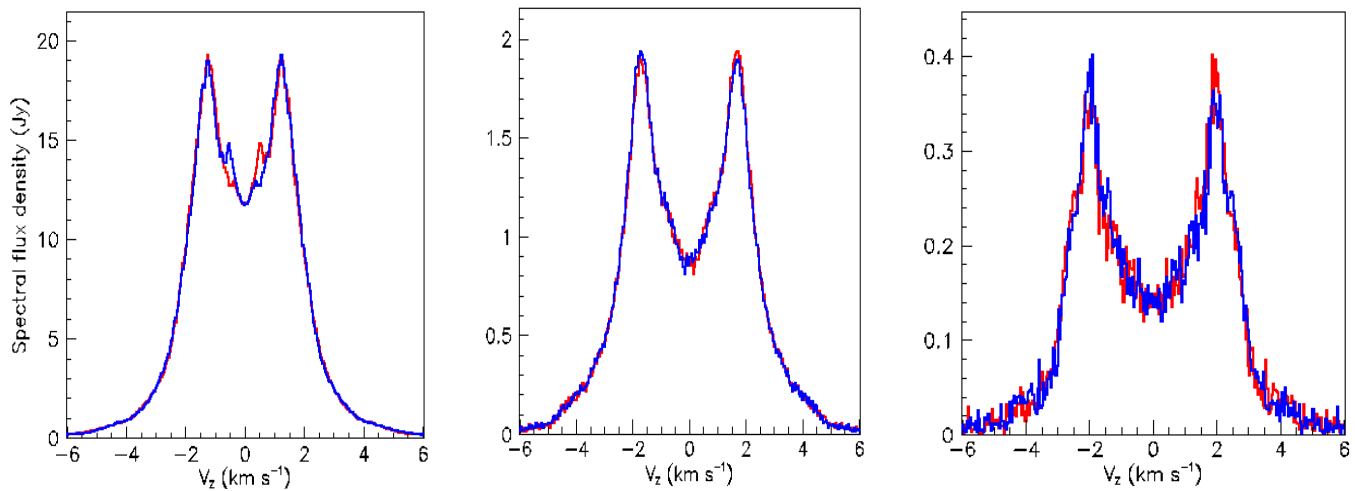

**Fig. 6.** From left to right: integrated spectra of CO(2-1), C$^{18}$O(2-1), and DCO$^{+}$(3-2) emissions integrated over the ellipses shown in Fig. 7. The blue histograms display spectra obtained by mirror symmetry about the origin.

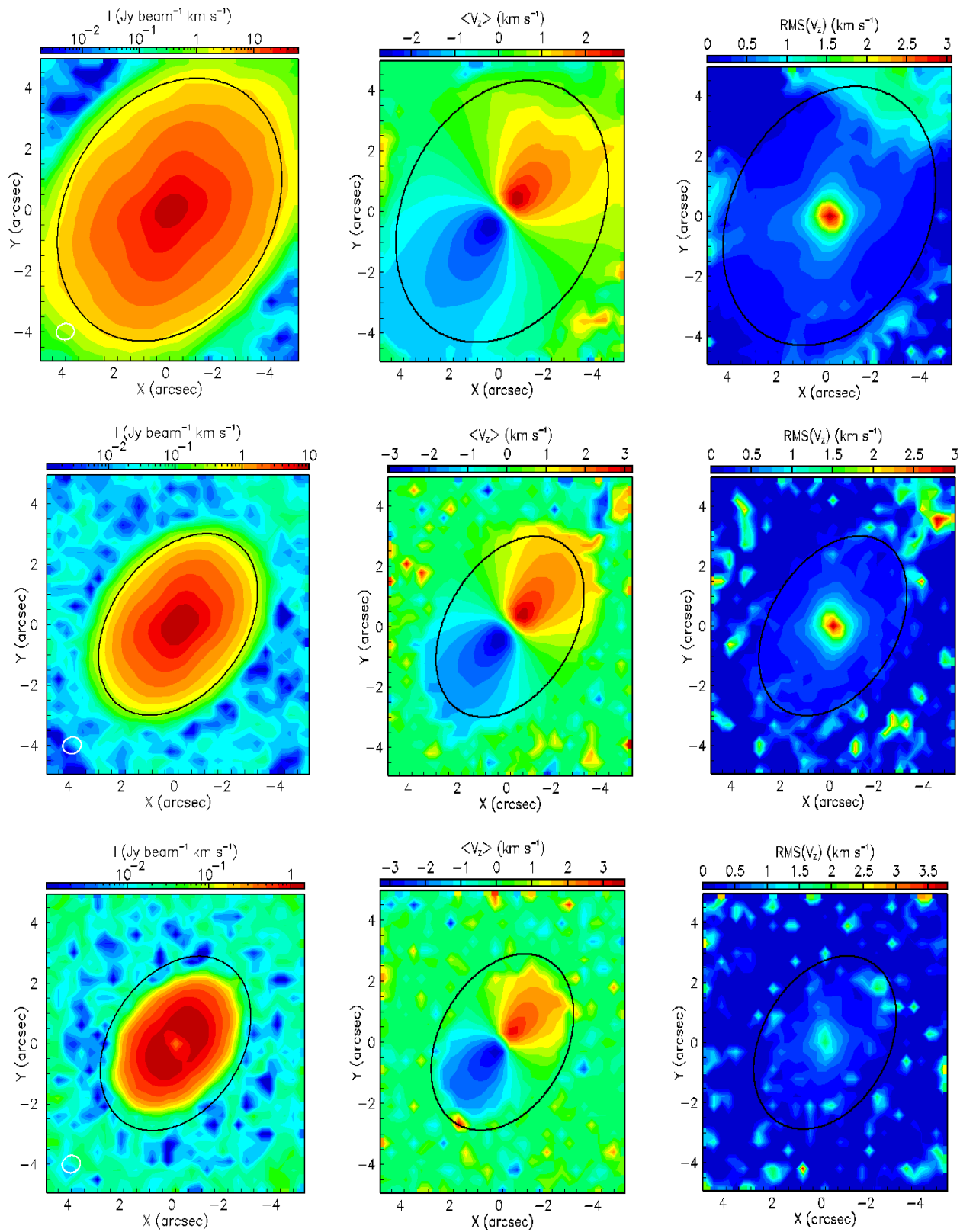

**Fig. 7. Sky maps of velocity integrated intensity (left), of $<V_z>$ (centre) and of Rms($V_z$) (right) for CO(2-1) (top panels), C$^{18}$O(2-1) (centre panels), and DCO$^+$(3-2) (bottom panels) emissions.** Ellipses indicate the regions of the sky plane generally retained in the analysis.

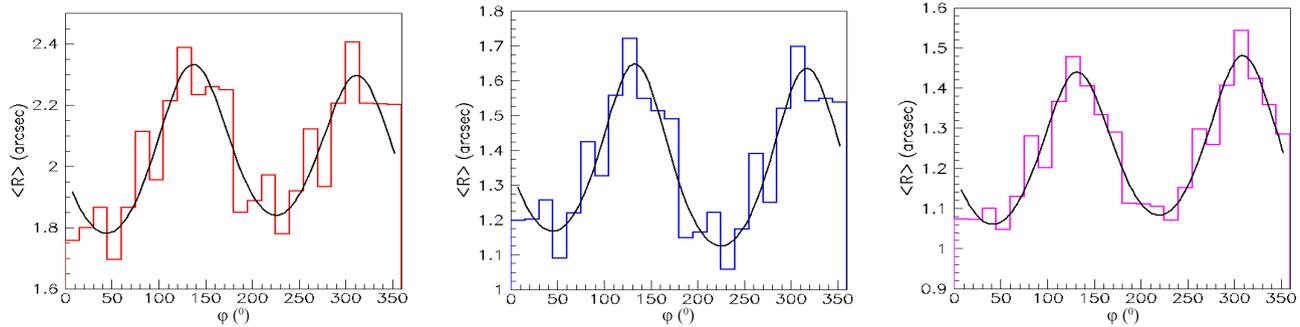

**Fig. 8. Dependence on $\varphi$ of <R> for CO (left), C$^{18}$O (centre), and DCO$^+$ (right).** The black curves are fits to ellipses (see text).

Repeating for the C$^{18}$O and DCO$^+$ lines the same geometry analysis that was performed for the continuum, we obtain semi-minor axes of 1.15 and 1.07 arcsec and semi-major axes of 1.64 and 1.46 arcsec, respectively. These correspond to inclinations with respect to the sky plane of 45$^0$ and 43$^0$, respectively (see Fig. 8), which is in excellent agreement with the continuum results. Similar but less accurate results are obtained for the optically thick CO(2-1) emission (39$^0$).

The position angles $\varphi_0$ are best measured from the maps of the mean Doppler velocity, which maximise the asymmetry between the blue-shifted and red-shifted components. The position angle obtained this way is 135$^0$ for each of the three lines, which is close to the continuum value of 130$^0$.

*Gas morphology*

Figure 9 shows maps of the de-projected intensity in the central region (both $|x'|$ and $|y'|$ smaller than 2 arcsec) and their projections on the x' and y' axes. The similarity between the x' and y' projections illustrates the validity of describing them in terms of inclined circular discs with nearly identical values of $\varphi_0$ and $\theta$. To a good approximation, they are both centred and symmetrical with respect to the origin, and the emission extends up to a distance from the central star of ~5.0 arcsec for CO(2-1), ~3.5 arcsec for C18O(2-1), and ~3.0 arcsec for DCO+(3-2).

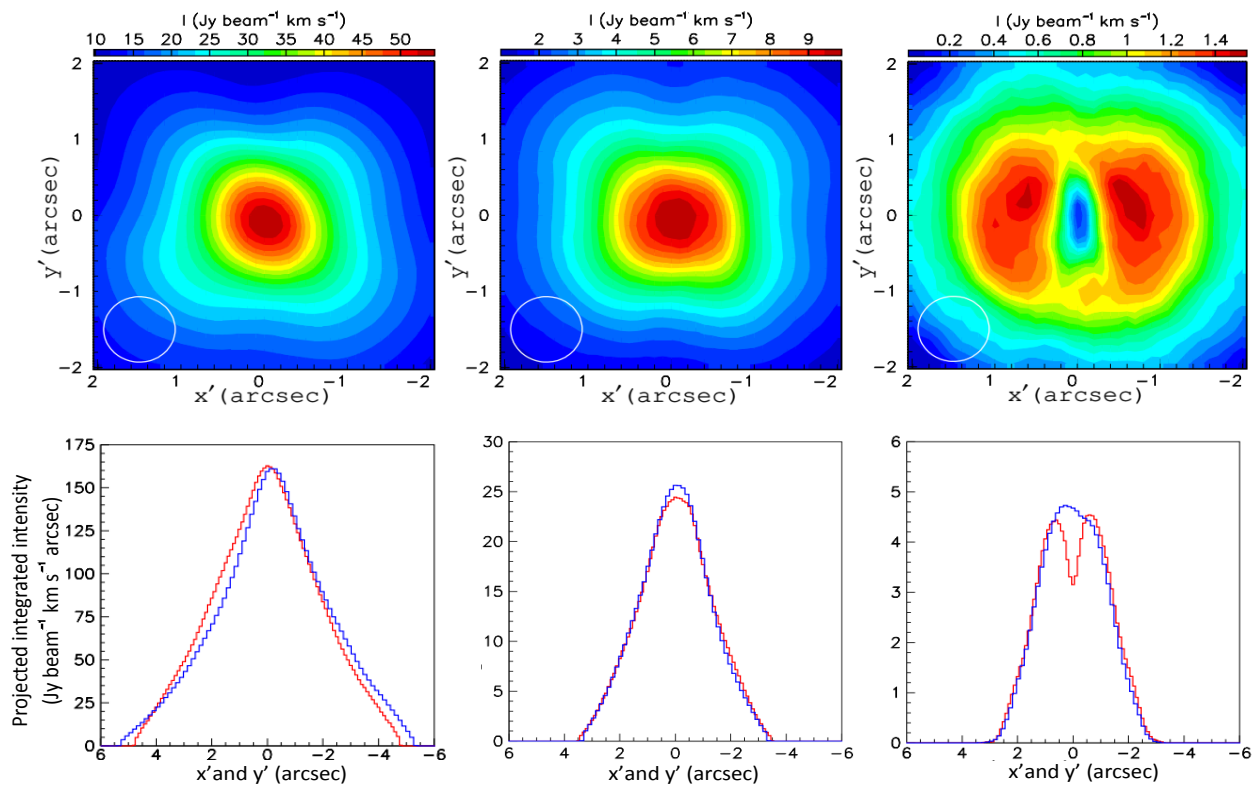

**Fig. 9. De-projected intensity of, from left to right: CO(2-1), C$^{18}$O(2-1), and DCO$^+$(3-2) emissions.** The top panels show the maps (y' vs x') and the bottom panels their projections (Jy beam$^{-1}$ km s$^{-1}$) on the x' (red) and y' (blue) axes.

Figure 10 (left) shows the dependence of intensity on $r'$ averaged over $\varphi'$. It displays decrement factors evaluated in the interval $1<r'<2$ arcsec of ~1.9 arcsec$^{-1}$ for CO(2-1), ~2.3 arcsec$^{-1}$ for C$^{18}$O(2-1), and ~2.7 arcsec$^{-1}$ for DCO$^+$(3-2) emissions, the latter dropping to small values near the star. As remarked by Isella, et al. [1], who observed a similar drop and noted that the DCO$^+$(3-2) and continuum intensities are similar at small distances from the star, an excessive continuum subtraction would wrongly cause a significant decrease of the line emission. However, the effect is too important to be an artefact of continuum subtraction. While both continuum subtraction and beam convolution contribute important uncertainties to the detailed morphology of the observed depression, its reality is not in doubt. Fig. 10 (right) displays the dependence on $\varphi'$ of the intensity averaged over $0.5<r'<2.0$ arcsec, which corresponds to the dust rings. Significant modulations are observed and are similar for the three lines, which is reminiscent of what was observed for the outer dust ring (Fig. 5 right). We emphasise that these modulations are unrelated to the disc inclination (we are dealing here with average intensity and not with average radius) but rather are associated with real intensity modulations in the disc plane.

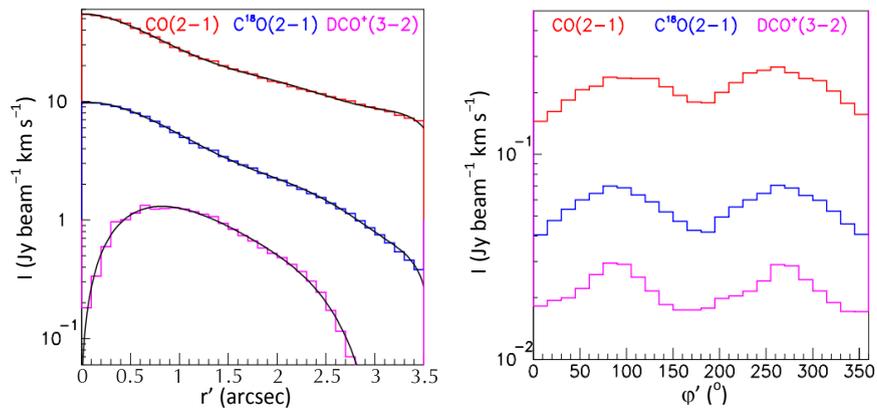

**Fig. 10. Dependence of the intensity on $r'$ (left) and $\varphi'$ (right, averaged over $0.5<r'<2$ arcsec).** Red is for CO(2-1), blue is for C$^{18}$O(2-1), and magenta for DCO$^+$(3-2).

The maps of the mean Doppler velocity in the disc plane are shown in Fig. 11. In contrast to the intensity, the Doppler velocity shows remarkably similar behaviour for each of the three lines (see next section).

Finally, to reveal the finer structure of the gas morphology, Fig. 12 shows the difference between the measured intensity and its mean value at the corresponding disc radius averaged over position angle. Fluctuations are observed at the ~10% level. As CO(2-1) emission is optically thick and the two other lines optically thin, these fluctuations probe the disc at different depths and indeed display different structures: a north-south bipolar asymmetry for CO emission corresponding to the disc surface and a north-south/east-west quadripolar asymmetry for C$^{18}$O(2-1) and DCO$^+$(3-2) probing the whole disc thickness.

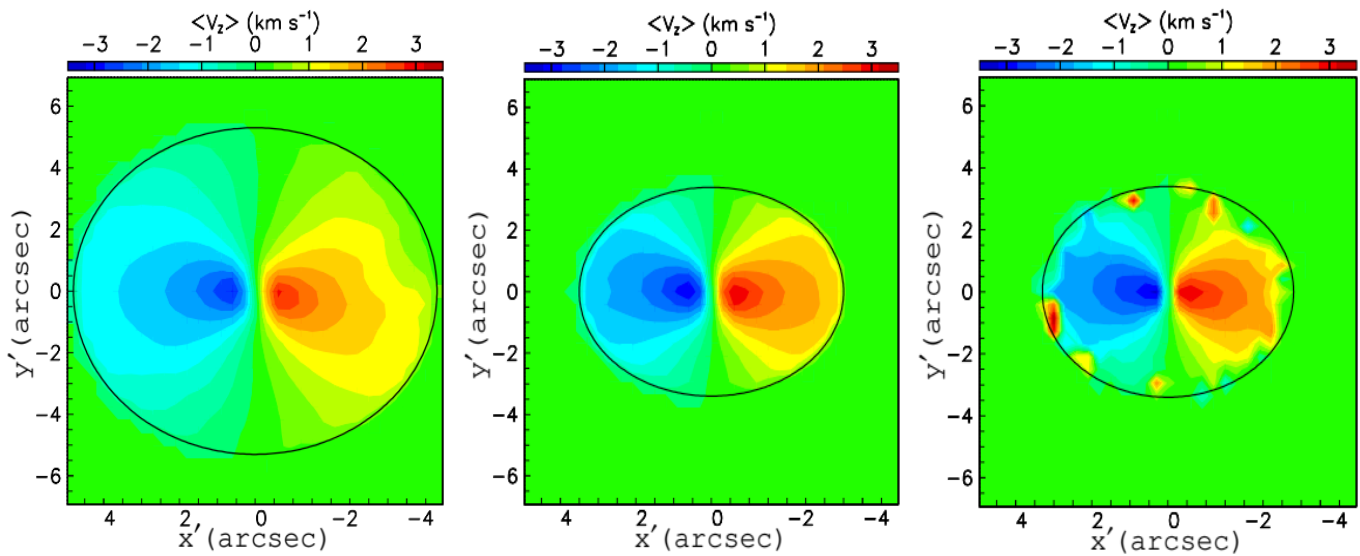

**Fig. 11. Disc plane maps of $<V_z>$ for CO(2-1) (left), C$^{18}$O(2-1) (centre), and DCO$^+$(3-2) (right).** Ellipses surround the regions where data are retained for the analysis.

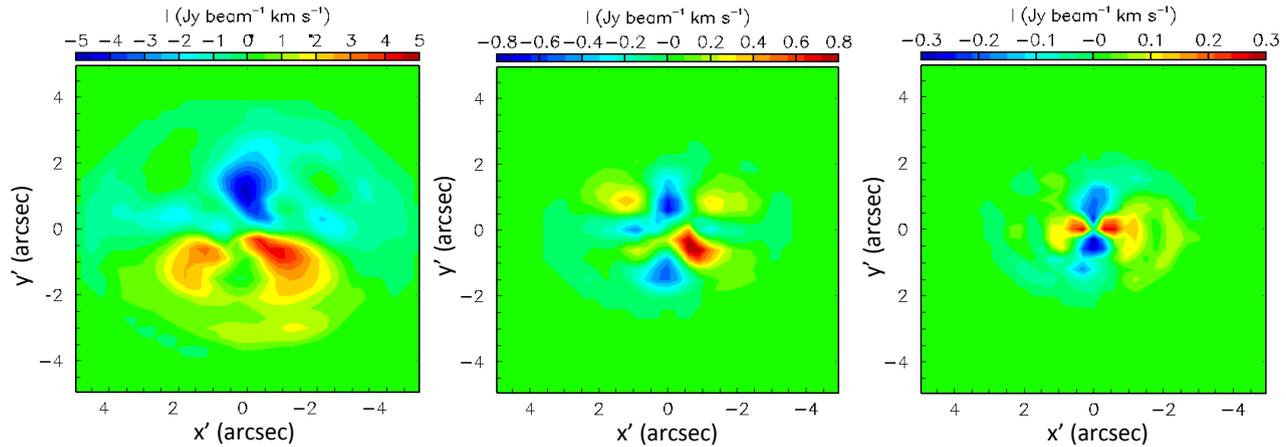

**Fig. 12. Intensity maps in the disc plane showing the difference between the measured intensity and its average at the same radius for each of CO(2-1) (left), C$^{18}$O(2-1) (centre), and DCO$^+$(3-2) (right).**

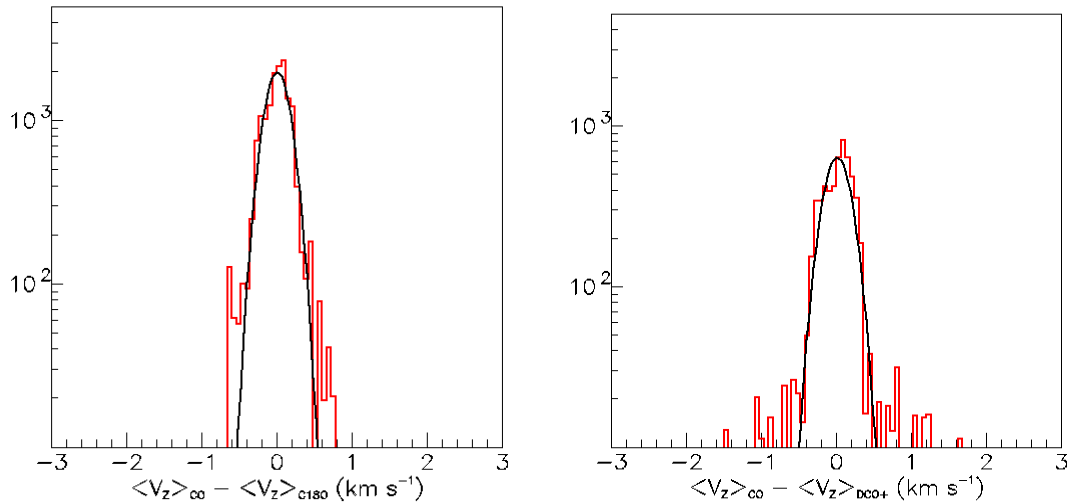

**Fig. 13. Distribution of the differences $<V_z>_{CO} - <V_z>_{C18O}$ (left) and $<V_z>_{CO} - <V_z>_{DCO+}$ (right) measured in the same pixel.** Each pixel contribution is weighted by the geometrical mean $(I_1 I_2)^{1/2}$ of the velocity integrated intensities $I_1$ and $I_2$ measured in the pixel for each of the two lines. The black curves show Gaussian fits to the peaks with σ's of 0.16 and 0.18 km s$^{-1}$ and mean values of 0.004 and 0.010 km s$^{-1}$, respectively.

*Gas kinematics*

As remarked in the preceding section, the three line emissions display nearly identical kinematics. This is illustrated in Fig. 13, which displays the distributions of the difference between the mean Doppler velocity measured in a given pixel for CO(2-1) and that measured in the same pixel for C$^{18}$O(2-1) (left) and DCO$^+$(3-2) (right). The mean values are 0.01 and 0.03 km s$^{-1}$ and the standard deviations 0.30 and 0.37 km s$^{-1}$, respectively.

In each datacube element, the Doppler velocity can be expressed as a function of rotation velocity, $V_{rot}$, and in-fall velocity, $V_{fall}$, as follows (Fig. 14):

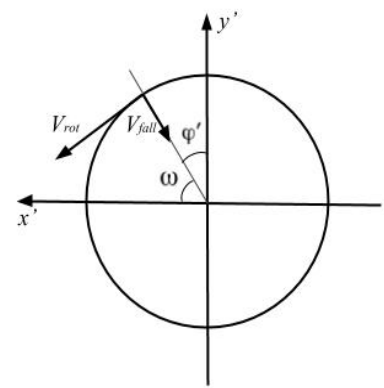

**Fig. 14. Velocity and trajectory of gas molecules in the disc plane.**

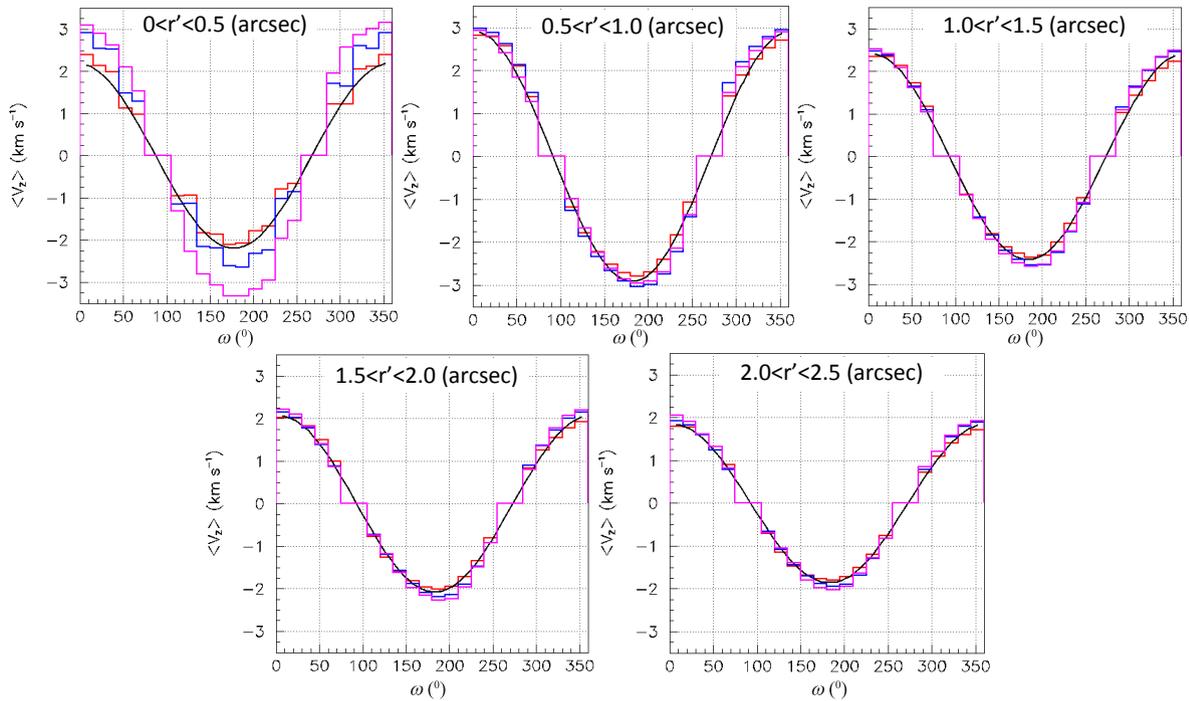

**Fig. 15. Dependence on $\omega=90°-\varphi'$ of $<V_z>$ for five intervals of $r'$ (see Table 3).** The black curves are fits of the form $V_0 \sin\theta \sin(\varphi'-\varphi'_0)$ to the CO(2-1) distributions. Red is for CO(2-1), blue for $C^{18}O(2-1)$, and magenta for $DCO^+(3-2)$.

$V_z = \sin\theta(V_{rot}\sin\varphi' - V_{fall}\cos\varphi') = V_0 \sin\theta \sin(\varphi'-\varphi'_0)$

with $V_{rot} = V_0 \cos\varphi'_0$ and $V_{fall} = V_0 \sin\varphi'_0$

namely $V_0 = (V_{rot}^2 + V_{fall}^2)^{1/2}$ and $\varphi'_0 = \tan^{-1}(V_{fall}/V_{rot})$.

Figure 15 shows the dependence of $<V_z>$ on $\varphi'$ averaged over five different intervals of $r'$ for each of the three lines separately. A fit of the form $<V_z> = V_0 \sin\theta \sin(\varphi'-\varphi'_0)$ is made to the CO(2-1) data in each interval of $r'$ separately. The small values obtained for $\varphi'_0$ show the dominance of rotation over in-fall. We estimate that the measured distributions can accommodate a maximal phase shift of ~5° with respect to pure rotation, implying an upper limit of ~9% on the ratio between in-fall and rotation velocities. The results are summarised in Table 3.

**Table 3. Parameters describing the gas kinematics.**

| CO(2-1) | | | | | |
|---|---|---|---|---|---|
| $r'$ interval | <0.5" | 0.5"-1.0" | 1.0"-1.5" | 1.5"-2" | >2.0" |
| $V_0$ (km s$^{-1}$) | 3.2 | 4.3 | 3.6 | 3.0 | 0.9 |
| $\varphi'_0$ | 1.3° | -0.5° | -2.4° | -2.5° | -2.9° |
| | CO(2-1) | $C^{18}O(2-1)$ | | $DCO^+(3-2)$ | |
| $r'$ interval | 0.8"-3.5" | 0.8"-3.5" | | 0.8"-3.0" | |
| $V_0^*$ | 4.1 | 4.4 | | 4.2 | |
| $n$ | 0.46 | 0.49 | | 0.40 | |

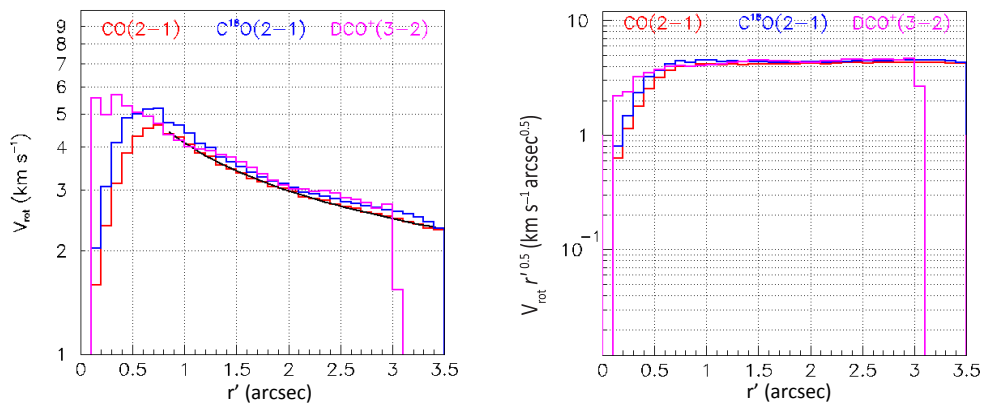

**Fig. 16. Dependence on $r'$ of $V_{rot}$ (left) and of $V_{rot} r'^{0.5}$ (right).** Red is for CO(2-1), blue for $C^{18}O(2-1)$, and magenta for $DCO^+(3-2)$.

Within the limits of a thin and flat disc and neglecting a possible in-fall velocity, the rotation velocity can be calculated for each data-cube element as $V_{rot}=V_z(\sin\theta\sin\varphi')^{-1}$ as long as $\sin\varphi'$ is not too small (in practice we require $|\sin\varphi'|>\sin 15°$). Fig. 16 (left) shows the dependence on $r'$ of $V_{rot}$ calculated in this manner. Fits of the form $V_{rot}=V^*_0 r'^{-n}$ yield the results listed in Table 3, providing evidence for approximate Keplerian motion. Clearer evidence is displayed in Fig. 16 (right) showing the dependence on $r'$ of the Kepler factor $V_{rot} r'^{0.5}$, which is constant for Keplerian rotation. It is indeed observed to be constant down to $r'\sim 0.8$ arcsec with a value of $\sim 4$ km s$^{-1}$ arcsec$^{1/2}$, in good agreement with the expectation of 3.9 km s$^{-1}$ arcsec$^{1/2}$ corresponding to a central mass of 2.3 solar masses.

*Line widths*

Figure 17 (left) shows the distributions of the difference between the Doppler velocity and its mean calculated in the same pixel, $\Delta V_z = V_z - <V_z>$, for each line separately (we recall that we require $|V_z|$ not to exceed 6 km s$^{-1}$). Gaussian fits to the central peaks (black curves) give standard deviations of 0.39, 0.32, and 0.31 km s$^{-1}$ for CO(2-1), C$^{18}$O(2-1), and DCO$^+$(3-2), respectively. The dependence on $r'$ and $\varphi'$ of $Rms(V_z)$ (averaged over $\varphi'$ and $r'$, respectively) is shown in Figs. 17 (right) and 18, the latter separately in the five intervals of $r'$. They display similar shapes for each of the three lines but different amplitudes, significantly larger for CO than for the two other emissions. Apart from beam convolution effects, the velocity dispersion is a combination of 1) instrumental resolution; 2) the range of velocities covered in each pixel because of disc thickness, disc inclination and Keplerian shear; 3) thermal broadening, having a standard deviation $\sigma_v=(2kT/\mu)^{1/2}$ where $k$ is Boltzmann constant, $T$ the gas temperature and $\mu = 28m_H$, $30m_H$, and $30m_H$ for CO, C$^{18}$O, and DCO$^+$, respectively; and 4) turbulence. All of these are expected to be significantly smaller than measured, never exceeding 0.1 km s$^{-1}$ for $r'>1$ arcsec. However, beam convolution, with FWHM values nearing 0.7 arcsec, seven times larger than the pixel size, is expected to dominate the line width. To evaluate its effect, we assume that the disc is thin and flat, that the orbits are circular, and that other contributions to the line width can be neglected. Moreover, we assume pure Keplerian motion with $V_{rot}r'^{1/2}=3.9$ km s$^{-1}$ arcsec$^{1/2}$, we use the $r'$ dependence of the intensity displayed in Fig. 10 (left), and we neglect its dependence on $\varphi'$. A same 3-$\sigma$ cut is applied to the model as to the data.

The results are compared with observations in Figs. 19, 20, and 21. The agreement is excellent considering the simplicity of the model. The apparent departure from Keplerian motion observed at distances from the star smaller than 0.8 arcsec is found to be largely due to beam convolution. The line width is reasonably well-reproduced by the model down to $r'\sim 0.5$ arcsec, where temperature and turbulence effects are likely to be more important. Above this distance, effects other than beam convolution do not significantly contribute to the line width.

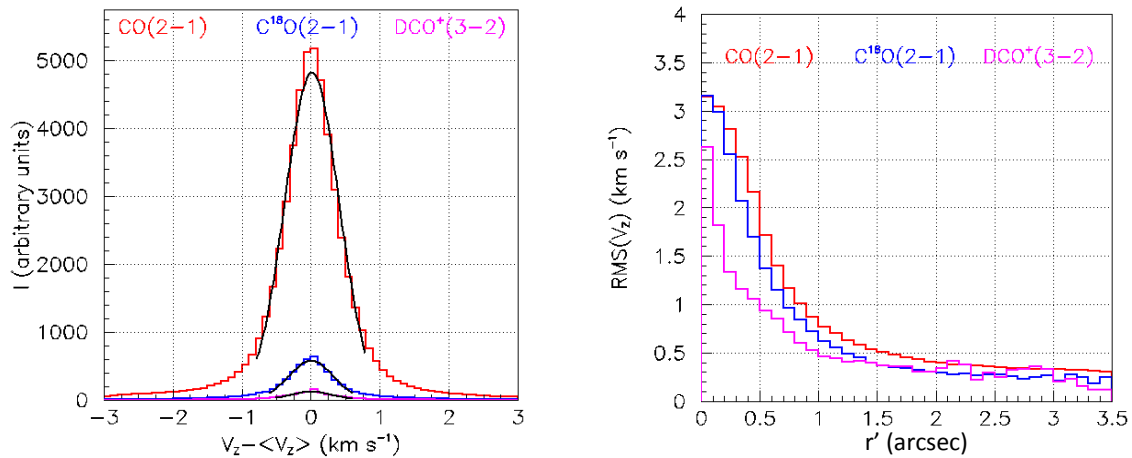

**Fig. 17.** Left: distribution of $V_z$-$<V_z>$ in each pixel and for each line separately. Right: dependence on $r'$ of $Rms(V_z)$. Red is for CO(2-1), blue for C$^{18}$O(2-1), and magenta for DCO$^+$(3-2).

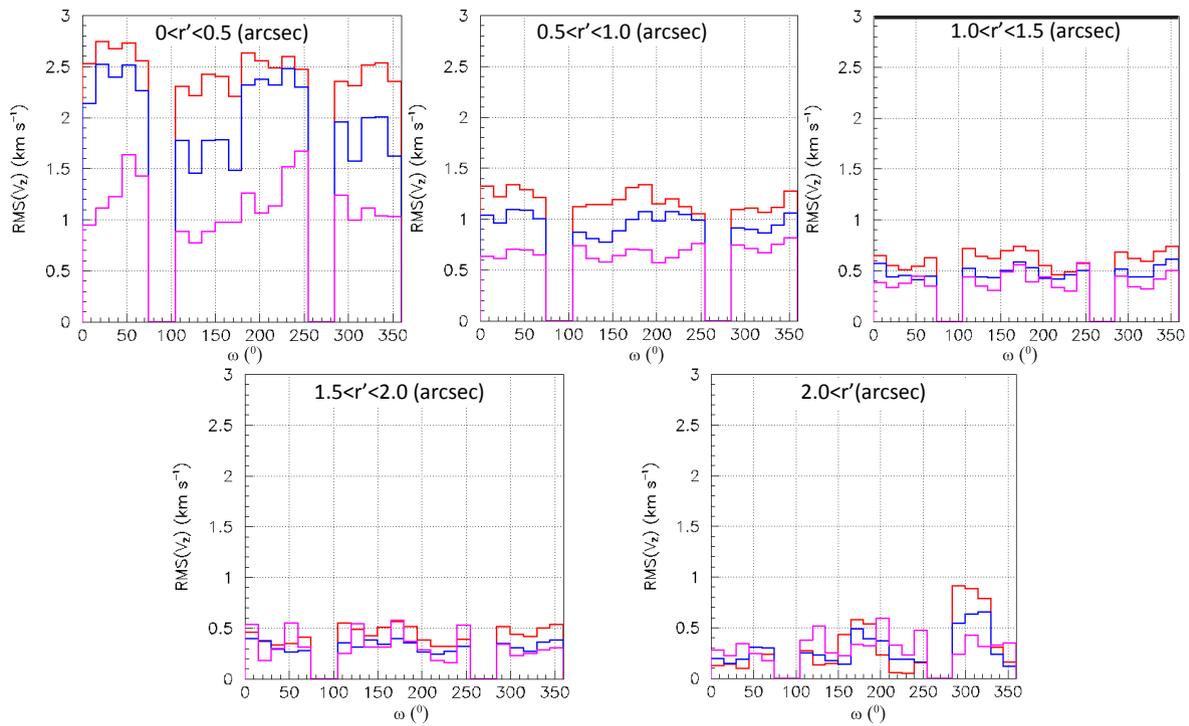

**Fig. 18. Dependence on $\omega=90°-\varphi'$ of Rms($V_z$).** Red is for CO(2-1), blue for $C^{18}O$(2-1), and magenta for $DCO^+$(3-2).

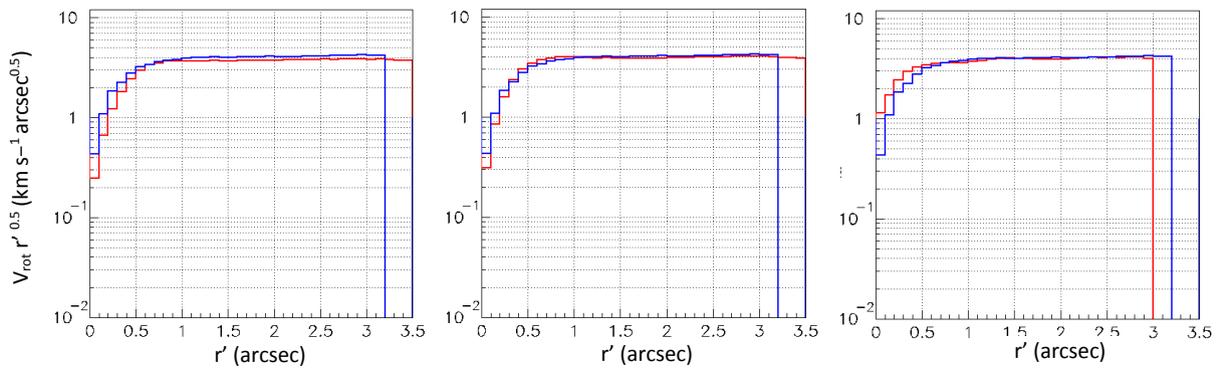

**Fig. 19. From left to right: radial dependence of the Kepler factor $V_{rot}r'^{0.5}$ as observed (red) and predicted by the model (blue) for CO(2-1), $C^{18}O$(2-1), and $DCO^+$(3-2), respectively.**

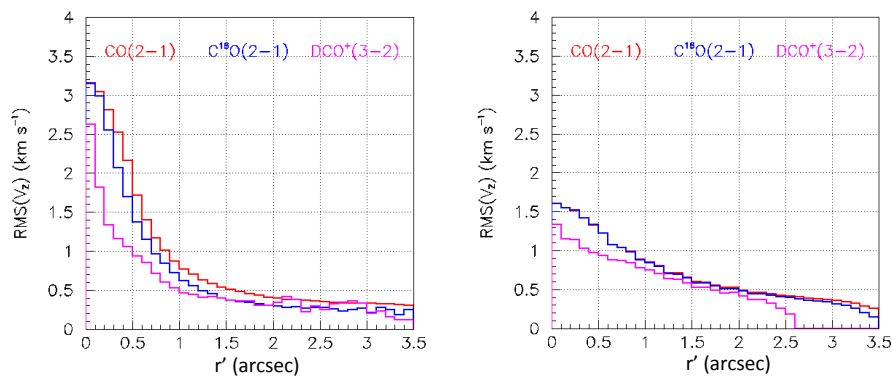

**Fig. 20. Radial dependence of Rms($V_z$) as observed (left) and predicted by the model (right).** Red is for CO(2-1), blue for $C^{18}O$(2-1), and magenta for $DCO^+$(3-2).

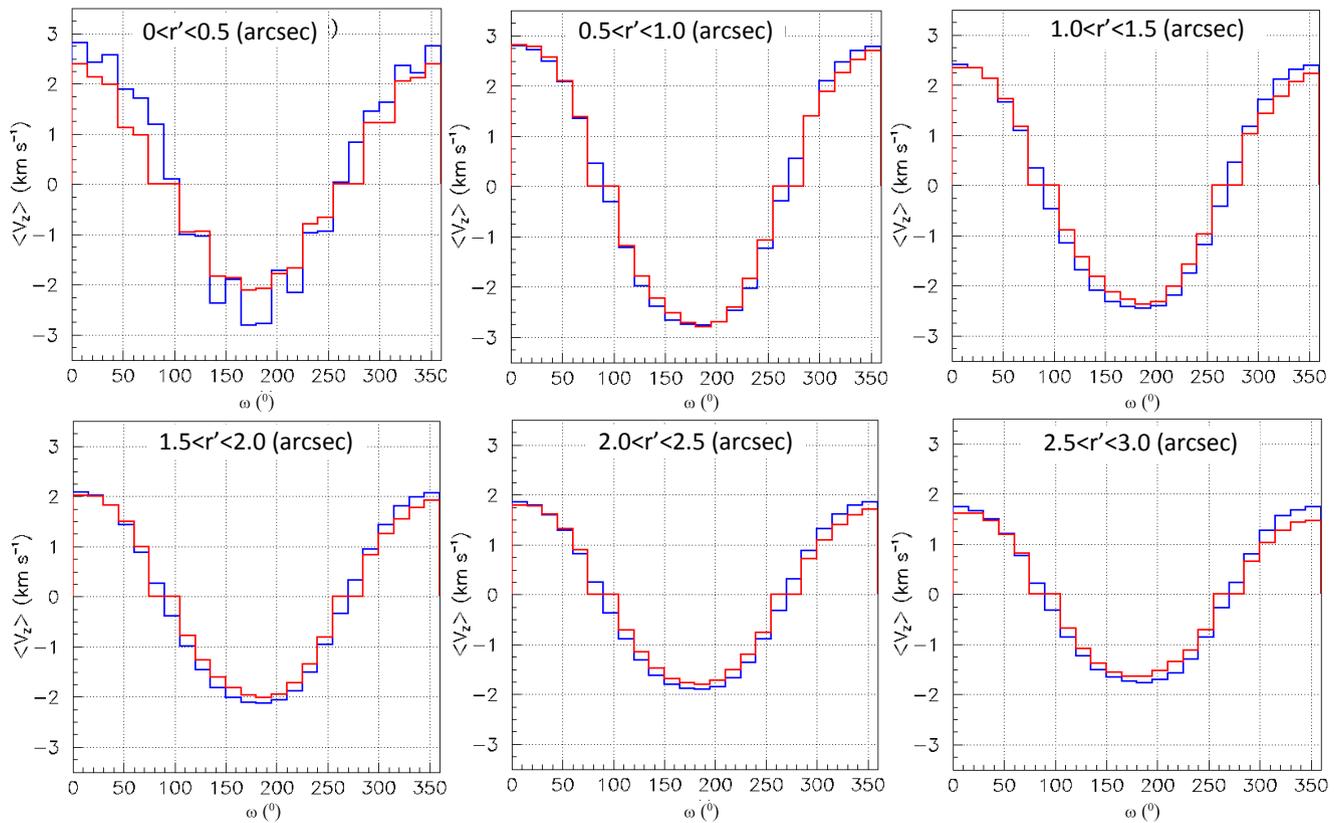

**Fig. 21. Dependence of $<V_z>$ (km s$^{-1}$) on $\omega=90°-\varphi'$ for six intervals of $r'$, 0.5 arcsec wide and covering between 0 and 3 arcsec.** Red is for the observations and blue for the model.

**Summary and conclusions**

Using ALMA observations of the continuum emission at 0.9 mm wavelength and of molecular line emissions from CO(2-1), C$^{18}$O(2-1), and DCO$^+$(3-2), the morphology of the dust and gas disc of HD 163296 has been explored; the position and inclination with respect to the sky plane have been accurately measured and found to be identical for gas and dust. The ring-gap structure of the dust disc has been described with better precision than previously achieved [1], and no evidence has been found for a third gap at a distance of ~1.31 arcsec from the star, as claimed by Isella, et al. [1]. The outer gap and ring have been shown to display a significant modulation of their morphology and intensity. The disc thickness at ~80 au has been shown to be smaller than 10 au.

Small intensity fluctuations in the three line emissions CO(2-1), C$^{18}$O(2-1), and DCO$^+$(3-2) have been revealed and precisely located and measured; they explore different regions across the disc thickness and display clearly different structures. In the region of the dust rings in particular, modulations of the intensity reminiscent of those displayed by the outer dust disc have been observed.

The three line emissions display remarkably similar kinematics dominated by Keplerian rotation. An upper limit of 9% of the rotation velocity has been placed on a possible in-fall velocity. A Keplerian factor of 4.0 km s$^{-1}$ arcsec$^{1/2}$ has been measured, consistent with the expectation of 3.9 km s$^{-1}$ arcsec$^{1/2}$ corresponding to a central mass of 2.3 solar masses. Apparent departure from Keplerian motion at distances from the star smaller than ~0.8 arcsec has been shown to be essentially due to beam size (~0.7 arcsec FWHM). The observed line widths have been shown to be dominated by beam size effects over most of the radial range. A simple model accounting for the effect of beam convolution has been shown to very well reproduce the observations.

Recently, several new protoplanetary discs displaying a ring-gap structure have been discovered and imaged with high resolution by ALMA (for a review, see Reference [11]). The complexity of their morphology, which includes spirals and clumps in addition to rings, has come as a surprise. This has triggered a surge of studies aimed at interpreting the gap-forming mechanism, usually in terms of planet formation (see for example Reference [12]). The present work has shown that the case of HD 163296 is particularly well-suited to give important contributions to this domain of stellar physics. The present data, which give evidence for strong radial depletion of the dust disc with no significant counterpart observed in the gas disc, do not strongly support the presence of planets in formation. However, higher resolution observations of the molecular line emissions should shed better light on this issue.

**ACKNOWLEDGEMENTS**

We thank Nguyen Xuan Que for contributing to this work in its earlier phase and Prof. P. Darriulat for guidance and advice. This paper makes use of data from the ALMA projects ADS/JAO.ALMA#2013.1.00601.S and ADS/JAO.ALMA#2013.1.00366.S. ALMA is a partnership of ESO (representing its member states), NSF (USA), and NINS (Japan), together with NRC (Canada), NSC, ASIAA (Taiwan), and KASI (Republic of Korea), in cooperation with the Republic of Chile. The ALMA Observatory is operated by ESO, AUI/NRAO, and NAOJ. We thank the ALMA staff for the reduction of the data and for their help with understanding the origin of a spurious emission in the CO(2-1) data at Doppler velocities well above the line emission. This research has made use of the SIMBAD database, operated at CDS, Strasbourg, France, and of the NASA ADS Abstract Services. The study is funded by the Vietnam National Foundation for Science and Technology Development (NAFOSTED) (No. 103.99-2016.50). We acknowledge support from the World Laboratory, Rencontres du Viet Nam, the Odon Vallet fellowships, the Vietnam National Space Center, the Graduate University of Science and Technology, and the Vietnam Academy of Science and Technology.

The authors declare that there is no conflict of interest regarding the publication of this article.